\documentstyle[12pt]{article}
\def\be{\begin{equation}}
\def\ee{\end{equation}}
\def\ba{\begin{eqnarray}}
\def\ea{\end{eqnarray}}
\def\la{~\mbox{\raisebox{-.6ex}{$\stackrel{<}{\sim}$}}~}

\def\bq{\begin{quote}}
\def\eq{\end{quote}}
\def\PL{{ \it Phys. Lett.} }
\def\PRL{{\it Phys. Rev. Lett.} }
\def\NP{{\it Nucl. Phys.} }
\def\PR{{\it Phys. Rev.} }

\font\tinynk=cmr6 at 10truept

\newcommand{\beq}{\begin{equation}}
\newcommand{\eeq}{\end{equation}}
\newcommand{\beqa}{\begin{eqnarray}}
\newcommand{\eeqa}{\end{eqnarray}}

\def\la{~\mbox{\raisebox{-.6ex}{$\stackrel{<}{\sim}$}}~}

\def\ltap{\ \raise.3ex\hbox{$<$\kern-.75em\lower1ex\hbox{$\sim$}}\ }
\def\gtap{\ \raise.3ex\hbox{$>$\kern-.75em\lower1ex\hbox{$\sim$}}\ }
\def\gl{\ \raise.5ex\hbox{$>$}\kern-.8em\lower.5ex\hbox{$<$}\ }
\def\roughly#1{\raise.3ex\hbox{$#1$\kern-.75em\lower1ex\hbox{$\sim$}}}

\begin{document}
\thispagestyle{empty}
\begin{flushright}
SU-ITP-99/52\\ hep-th/9912125\\ December 1999
\end{flushright}
\vspace*{1cm}
\begin{center}
{\Large \bf Crystal Manyfold Universes in $AdS$ Space}\\
\vspace*{1.5cm}
Nemanja Kaloper \\
\vspace*{0.2cm}
{\it Department of Physics} \\
{\it Stanford University}\\
{\it Stanford, CA 94305-4060}\\
{\tinynk kaloper@epic.stanford.edu}\\
\vspace{1.5cm}
ABSTRACT
\end{center}
We derive crystal braneworld solutions, comprising of
intersecting families of parallel $n+2$-branes
in a $4+n$-dimensional $AdS$ space. Each family
consists of alternating positive and negative tension branes.
In the simplest case of exactly orthogonal families, there
arise different crystals with unbroken $4D$ Poincare
invariance on the intersections, where our world can reside.
A crystal can be finite along some direction, either because that
direction is compact, or because it ends on a segment of $AdS$ bulk,
or infinite, where the branes continue forever.
If the crystal is interlaced by connected
$3$-branes directed both along the intersections and
orthogonal to them, it can be viewed as
an example of a Manyfold universe proposed recently by
Arkani-Hamed, Dimopoulos, Dvali and the author.
There are new ways for generating hierarchies, since the
bulk volume of the crystal and the lattice spacing affect the
$4D$ Planck mass.
The low energy physics is sensitive to the boundary conditions
in the bulk, and has to satisfy the same constraints
discussed in the Manyfold universe. Phenomenological considerations
favor either finite crystals, or crystals which
are infinite but have broken translational invariance in the bulk.
The most distinctive signature of the
bulk structure is that the bulk gravitons are Bloch waves,
with a band spectrum, which we explicitly construct
in the case of a $5$-dimensional theory.

\vfill
\setcounter{page}{0}
\setcounter{footnote}{0}
\newpage

Recently a remarkable observation
by Arkani-Hamed, Dimopoulos and Dvali
that there may exist additional
sub-millimeter spatial dimensions \cite{add} has generated
tremendous interest. In such theories, the usual Standard Model
degrees of freedom are localized to a $3$-brane which is embedded
in a higher-dimensional bulk, and thus at low energies are
indifferent to the extra dimensions. On the other hand, gravity
and, typically, other weakly coupled fields live in the bulk,
probing equally all spatial dimensions. The observed
weakness of gravity in four dimensions is generated naturally,
because by spreading through all spatial dimensions gravity becomes
softer, as can be seen immediately from Gauss law.
Hence the hierarchy between the Planck scale, $M_{Pl} \sim 10^{19} GeV$
and the electroweak scale, $m_{EW} \sim TeV$ and its radiative stability,
can be naturally explained if the size of the $n$
internal dimensions $r$ is large:
$M_{Pl}^2 = M^{2+n}_* r^n \gg M^2_* \sim m^2_{EW}$ \cite{add}.

Studies of phenomenological constraints in
\cite{add} have confirmed that such models are consistent
with observations, with
the unification scale as low as ${\rm few} \times 10~TeV$.
There has subsequently been much interest in
models with large extra dimensions
\cite{aadd}-\cite{astconstr}.
String theory may give rise to such models, with the
radius of compactification only a few orders of magnitude
above the Planck scale, such as in Ho\v rava-Witten \cite{hw,low},
intermediate sizes $\sim (TeV)^{-1}$ \cite{ia,late}, and even
$\sim (M_{Pl}/M)^{2/n} M^{-1} \la mm$ \cite{aadd,lstr}.

The concept of localization of matter to the branes plays a
key role in masking the large extra dimensions from observation at
low energies. In fact, the universe as a domain wall in a non-compact
space has been considered before \cite{rusha,old}, but there
gravity was higher-dimensional at all scales. But if gravity itself is
localized to the 3-brane, then $4D$ gravity can be reproduced at low energies
even if the extra dimensions are much larger than a $mm$. A very interesting
recent construction by Randall and Sundrum \cite{rs} provides an
elegant example of this idea for the case of one extra dimension,
which is linearly infinite, but has finite proper volume.
Similar ideas have been explored also in \cite{gog}.
Subsequently it has been shown that gravity can be localized
to intersections of $n+2$ branes in $4+n$-dimensional spaces
\cite{addk1}. More examples were found later \cite{csn},
and various
issues \cite{theoads}-\cite{cosads} were considered.
Since the Randall-Sundrum proposal \cite{rs} employs $AdS$
bulk geometry, it is possible to view the model
in the framework of the $AdS/CFT$ correspondence \cite{malda,adscft}.
Interpreting the fifth dimension as the scale of the $4D$
theory \cite{lenny} and the evolution along it as the holographic
renormalization group \cite{hrg}, this model can be rephrased
as a $4D$ $CFT$ theory coupled with gravity \cite{gubs}, without
invoking extra dimensions. The $AdS/CFT$ correspondence has also
been pursued in \cite{verlinde}.

On a different track, it is possible that our $3$-brane is not
alone in the bulk. In fact, the bulk may be populated by many
other branes, if we wish to understand the weak breaking of
symmetries by shining in the bulk \cite{shining}. Hence, the
universe may contain many nearby branes, which are dark because
the light travels only along them, but are felt by their
gravity \cite{addk2}. This Manyfold universe
has been considered recently by
Arkani-Hamed, Dimopoulos, Dvali and the author \cite{addk2},
and it has been shown that
it can be consistent with the observations, while giving rise
to many new phenomena: dark matter candidates which are dissipationful,
dark stars, hybrids, neutrino mixing, SUSY breaking et cetera.
Some solutions representing multi-brane configurations in $5D$
spacetimes have been constructed \cite{multib}, and some aspects of brane
networks considered \cite{nets,nam}.

In this paper, we will construct explicit solutions of Einstein's
equations which represent intersecting families of parallel
$n+2$-branes in $4+n$-dimensional $AdS$ spacetimes. Each family
will contain branes of alternating tension. While we are a little
reluctant to treat the negative tension branes on equal footing
with the positive tension ones, we will ignore this when
considering the solutions because of considerable technical
simplification. If the bulk cosmological constant is replaced by
a potential, there may exist solutions which involve only positive
tension branes, while sharing some generic properties with our
solutions. Alternatively, a first-principles construction of branes
with negative tension may yet ensue. We will show that there arise
several different crystal solutions, all of which support the
usual $4D$ gravity. However phenomenological considerations
strongly favor finite crystals or crystals which are infinite but
have strongly broken translational symmetry, where our world
arises as a lattice defect. If a crystal is interlaced by a
grid of $3$-branes, some along the intersections and others
stretching between them, the low energy physics of such a solution
is identical to that of a Manyfold universe \cite{addk2}. Such a crystal
would have to be subject to the same astrophysical constraints as
any generic Manyfold \cite{addk2}. We will
further show that the spectrum of bulk gravitons has band
structure, with a gap at zero energy, because the bulk gravitons
are Bloch waves, like electrons in ordinary crystals. This may
allow the string scale in a $5D$ theory to be as low as $M_* \sim
100 TeV$, at the cost of having ${\cal N} \sim 10^{16}$ branes
in the $5D$ $AdS$ spacetime. However, the reward consists of
reproducing the $4D$ Planck scale $M_{Pl} \sim 10^{19} GeV$, the
electroweak scale $m_{EW} \sim 1 TeV$ and suppressing the
corrections to Newton's law at distances $\ge 1 mm$. It may be
possible to reduce the number of required branes by going to more
than one extra dimensions.

We begin by deriving the solution describing the crystal braneworlds.
Consider an array of $n$ orthogonal families of
$n+2$-spatial dimensional branes in a $4+n$-dimensional spacetime,
with a bulk cosmological constant $\Lambda$.
For simplicity we take the branes in each family
to have identical tension $\sigma_k$.
The field equations can be
derived from the action
\be
S = \int_M d^{4+n} x \sqrt{g_{4+n}}
\Bigl(\frac{R}{2\kappa^2_{4+n}} + \Lambda\Bigr)
 - \sum_{k=1}^n \sum_{j_k} \int_{j_k l_k} d^{3+n} x \sqrt{g_{3+n}} \sigma_k .
\label{action}
\ee
Here $\kappa^2_{4+n} = 8\pi/M^{n+2}_*$, where $M_*$ is the
fundamental scale of the theory. In the brane actions, $g_4$
refers to the induced metric on the brane.
Note that the measure of integration differs between each brane,
and between the branes and the bulk. This will be reflected
in the field equations, where ratios
$\frac{\sqrt{g_{3+n}}}{\sqrt{g_{4+n}}}$
weigh the $\delta$-function sources. The gravitational field equations are
\be
G^a{}_b = \kappa^2_{4+n} \Lambda \delta^a{}_b - \sum_{k=1}^n
\frac{\sqrt{g_{3+n}}}{\sqrt{g_{4+n}}}|_k \kappa^2_{4+n} \sigma_k
\sum_{j} (-1)^{j} \delta(z^k - j l_k) {\rm diag}
(1,1,1,1,1,...,0_k,...,1)
\label{eoms}
\ee
where the coordinates
${ z}^k$ parameterize the extra dimensions. In general, these
equations should be supplemented by the equations of motion for
the branes, which however are solved by placing branes at fixed
locations parallel to the $AdS$ boundary. The issue of stability
would then have to be addressed, and is equivalent to providing
the brane Goldstone bosons with mass terms. We will not address
this in detail here. The ratios
$\frac{\sqrt{g_{3+n}(k)}}{\sqrt{g_{4+n}}}$ which appear in
(\ref{eoms}) reduce to $\sqrt{g^{kk}}$ for simple diagonal
metrics, however in general they cannot be gauged away.

Deriving the solutions is straightforward \cite{addk1}.
Away from the branes, the bulk geometry consists of patches of
the $4+n$-dimensional $AdS$ space.
On the branes, there are additional $\delta$-function
singularities, which cause discontinuities in the derivatives of
the metric. To find the global solution, valid on the complete
manifold, we should seek the solution in the form
\be
ds^2_{n+4} = \Omega^2
\Bigl(\eta_{\mu\nu} dx^\mu dx^\nu + \sum^n_{k=1} (d{ z}^k)^2 \Bigr)
\label{ansatz}
\ee
and adjust the boundary conditions for $\Omega^{-1}$ such that it
solves Einstein's equations (\ref{eoms}), and between the branes
reproduces the metric of $AdS$ bulk. It is convenient to
recast the Einstein's equations (\ref{eoms}) in the conformal
frame $\tilde g_{ab} = \Omega^{-2} g_{ab}$. We use
the standard relation for $4+n$ spacetime dimensions
\ba
G_{ab} &=& \tilde{G}_{ab} + (n+2) \left( \tilde{\nabla}_a \mbox{log}
\Omega
\tilde{\nabla}_b \mbox{log} \Omega - \tilde{\nabla}_a \tilde{\nabla}_b
\mbox{log} \Omega\right) \nonumber \\
&&\,~~~~
+ (n+2)\tilde{g}_{ab} \left(\tilde{\nabla}^2 \mbox{log} \Omega
+ \frac{n+1}{2} (\tilde{\nabla} \mbox{log} \Omega)^2 \right).
\label{conformal}
\ea
Since the conformal metric is flat, $\tilde G_{ab} = 0$. Then
substituting (\ref{conformal}) into (\ref{eoms}) yields two
equations for $\Omega^{-1}$:
\ba
\vec \nabla^2 \Omega^{-1} &=& \frac{\kappa^2_{4+n} \sigma_k}{(n+2)}
\sum_{k=1}^n \sum_{j}
(-1)^{j}
\delta(z^k - jl_k), \nonumber \\
(\vec \nabla \Omega^{-1})^2 &=& 2
\frac{\kappa^2_{4+n} \Lambda}{(n+2)(n+3)}.
\label{omegaeqs}
\ea
Note that in deriving the first equation it was important to use
$\frac{\sqrt{g_{3+n}}}{\sqrt{g_{4+n}}} = \Omega^{-1}$, which
ensures that the brane tensions are constant.

It is now clear from the first of equations (\ref{omegaeqs}) that
$\Omega^{-1}$ is a linear combination of Green's functions for
each family of parallel branes. Since the tensions $\sigma_k$ are
the strengths of sources, they control the discontinuity of the
derivatives of $\Omega^{-1}$ on each brane. It is easy to see that
the discontinuities must satisfy $(\partial_k \Omega^{-1})^2 =
\frac{\kappa^4_{4+n}\sigma^2_k}{4(n+2)^2}$. Thus the discontinuity
in the gradient of $\Omega^{-1}$ on each intersection of
branes must satisfy
$(\vec \nabla \Omega^{-1})^2 = \frac{n
\kappa^4_{4+n}\sigma^2_k}{4(n+2)^2}$.
But because the second of eqs. (\ref{omegaeqs}) is defined
globally, these two expressions must coincide, leading to
\be
\kappa^4_{4+n} \sigma^2_k = \frac{8(n+2)}{n(n+3)}
\kappa^2_{4+n} \Lambda.
\label{finetun}
\ee
Hence the condition for the existence of the solutions is to
fine tune the tensions of the branes to the bulk cosmological
constant according to (\ref{finetun}). Then as we see from
(\ref{ansatz}) the induced metric on the intersections
is Minkowski, meaning that the conditions (\ref{finetun})
amount to fine tuning the effective $4D$ cosmological constant
on each intersection to zero.

We can now solve the first of (\ref{omegaeqs}) as follows.
Suppose that in each family there is ${\cal N}_k$
parallel branes, which for reasons of simplicity we have taken to be
equidistant, separated by $l_k$, as is seen in eqs.
(\ref{eoms}) and (\ref{omegaeqs}). The solution
can be easily generalized when the distances between the branes
vary within each family, as will become clear from the construction
below. The eqs. (\ref{omegaeqs}) yield the solution for
the conformal factor $\Omega$
\be
\Omega^{-1} = K \sum_{k=1}^n {\cal S}(z^k) + 1
\label{omsol}
\ee
where $K = (\sqrt{n}L)^{-1}$, and the functions ${\cal S}(z^k)$
satisfy
\ba
&& \frac{d^2 {\cal S}(z^k)}{d(z^k)^2} = 2 \sum_{j} (-1)^{j}
\delta(z^k - jl_k),
\nonumber \\
&&
|\frac{d{\cal S}(z^k)}{dz^k}| = 1.
\label{tsfun}
\ea
These two equations follow from (\ref{omegaeqs}) by using linear
superposition, eq. (\ref{finetun}) and the definition of the
$AdS$ radius. The integration constant in (\ref{omsol})
must be nonzero in order to excise the $AdS$ boundary from the
manifold, and is set to unity
by gauge choice.
As a result, the warp factor never diverges; otherwise
there could not exist a normalizable $4D$ graviton mode.

It is easy to see that the function ${\cal S}(z^k)$ which solves
eqs. (\ref{tsfun}) must be the sawtooth function.
We can write it as follows:
\be
{\cal S}(z^k) = \cases{ ...\cr
2pl_k - z^k, & {\rm for} ~$(2p-1) l_k < z^k < 2p l_k$;\cr
z^k-2pl_k, & ~~~~~ $2pl_k < z^k < (2p+1) l_k$; \cr
...~~.}
\label{sawt}
\ee
The solutions differ globally depending on the boundary conditions
for each ${\cal S}(z^k)$. In general there are four distinct
types of boundary conditions:

$(i)$ There is a discrete infinity of branes along the $k^{\rm th}$
axis, and the crystal is infinite in this direction;

$(ii)$ There is ${\cal N}_k = 2N_k + 1$ branes, and
$N_k+1$ have positive tension and $N_k$ negative; the crystal ends on
branes with positive tension in each direction along the $z^k$
axis; outside of the crystal there is the near-horizon geometry of
$AdS$ space, with the horizon located at infinite proper spatial
distance from each end;

$(iii)$ The crystal is semi-infinite along the $z^k$ axis,
with infinitely many branes, and boundary conditions satisfying
$(i)$ on one end and $(ii)$ on the other; the brane at one end
must have positive tension;

$(iv)$ The crystal is finite along $z^k$, with an
even number ${\cal N}_k = 2N_k$ of distinct branes along the $z^k$ axis,
a half of them with positive and a half with negative tension; the
solution can be viewed as case $(ii)$ but with the $0^{\rm th}$
and $2N_k+1^{\rm th}$ branes identified; this means, that the
$z^k$ axis is compactified on a circle.

By linear superposition, the global solution for the crystal is
a combination of any of the four possibilities realized along
individual axes. For the case $(i)$, the solution for the sawtooth
functions is adequately represented by an infinite array of the
form (\ref{sawt}). In the other three cases, by choosing the origin
of the coordinate system in the bulk to lie at the end
brane with positive tension, we can rewrite the solution in a more
compact form:
\be
{\cal S}(z^k) = 2\sum_{j=0}^{{\cal N}_k} (-1)^j \theta(z^k - j l_k)
(z^k - jl_k) -  z^k,
\label{sawtcom}
\ee
where ${\cal N}_k = 2N_k + 1$ in the case $(ii)$,
${\cal N}_k \rightarrow \infty$ in the case $(iii)$,
and ${\cal N}_k = 2 N_k$ in the case $(iv)$.
Therefore, in terms of these functions, the metric can be written
down as
\be
ds^2_{4+n} = \frac{1}{(K \sum_{k=1}^n {\cal S}(z^k) + 1)^2} \Bigl(
\eta_{\mu\nu} dx^\mu dx^\nu + \sum_{j=1}^n (dz^j)^2 \Bigr).
\label{finsol}
\ee
This is the solution for the crystal Manyfold
universe. By the conditions (\ref{finetun}), all the vertices of the
crystal lattice correspond to the $4D$ Minkowski spacetimes, and can
accommodate $3$-branes which localize the Standard Model
particles. Therefore, our world could reside at any one of the
vertices, leading to different low energy theory, as long as there
exists a massless $4D$ graviton mode. The $3$-branes lying on the
other vertices would behave as mirror worlds, and can be connected
to our own by additional $3$-branes embedded in the intersecting
$n+2$-branes. Hence a crystal interlaced by $3$-branes would give
rise to a Manyfold universe structure of \cite{addk2}.

It is straightforward to verify that the theory
defined by the action (\ref{action})
in general admits $4D$ massless graviton modes on the backgrounds
(\ref{finsol}). Hence at large distances these modes will give
leading contributions to the gravitational interactions between
particles localized to the intersections, reproducing the usual
Newton's law and masking the extra dimensions.
The wave function of these modes is
\be
h_{\mu\nu~0}(x^\mu,\vec z) \sim \Omega^{(2-n)/2}(\vec z)
e^{ik_\mu x^\mu},
\label{zeromode}
\ee
as we will see below in more detail. Using (\ref{action}) and
(\ref{zeromode}), by ignoring the massive KK modes and reducing
the theory (\ref{action}) to four dimensions, we find that
the $4D$ Planck mass is
\be
M^2_{Pl} = M^{n+2}_* \int_{\rm crystal~space} d^n \vec z ~\Omega^{n+2}.
\ee
The integral has to be performed over the whole crystal
space, which is composed
of a number of patches of a different kind. Depending on the type,
there may be semi-infinite patches comprised of the near-horizon
segments of $AdS$ bounded by branes of positive tensions.
Each such patch is identical to the patches considered in the case of
a single intersection of branes, contributing
$\frac{1}{(n+1)! K^n} M^{n+2}_*$ to $M_{Pl}^2$ \cite{addk1}.
However, since now there are also patches of geometry which
are completely bounded by branes, there arise additional
contributions to $M_{Pl}^2$. Different patches, or cells,
can be classified by counting how many branes they are bounded by,
such that the integrations in those directions would
be bounded between $0$ and $l_k$.
It is clear from (\ref{omsol}) and (\ref{sawt})
that all topologically identical segments
give the same contribution to the mass integral.
Each distinct integral needs to be
weighed by the symmetry factor, which counts the total number of cells of
a given class in the crystal. The integrals
$I = \int_{\rm cell} d^n \vec z ~\Omega^{n+2}$ in general assume the
form
\be
I = \int^{l_n}_0 dz^n
... \int^{l_{p+1}}_0 dz^{p+1} \int^{\infty}_0 dz^p ... \int^{\infty}_0 dz^1
\frac{1}{(K \sum_{j=1}^n z^j + 1)^{n+2}},
\label{int}
\ee
where $p$ is the number of directions which end on the
$AdS$ horizon.
The integral can be evaluated straightforwardly,
giving
\be
I = \frac1{(n+1)! K^n} \sum_{j=0}^{n-p} (-1)^j
\sum_{{\cal B}_j} \frac{1}{(1+K \sum_{i \in {\cal B}_j} l_i)^2},
\label{intev}
\ee
where ${\cal B}_j$ are combinations of $\{l_{p+1},...,l_n\}$
of length $j$.
To compute the symmetry factors, we break up
the crystal into topologically inequivalent cells.
These are distinguished by how many of their sides end on
$AdS$ horizon. Clearly, the cells inside the crystal only end on branes,
and correspond to taking $p=0$ in (\ref{intev}).
We can count the cells in each topological class
as follows. In the case of $1D$ crystals, we can
divide the cells along the crystal into
two types, ${\cal A}$ and ${\cal B}$. The type ${\cal A}$
are the cells inside the crystal, which end on branes on both
sides, and there is ${\cal N}_k$ of them. The type ${\cal B}$ cells
end on the $AdS$ horizon, and there is at most two of them.
Then the crystal (\ref{finsol}) can be represented formally
as a direct product
\be
{\rm Brane~ Crystal} \equiv \prod_{\bigotimes ~j=1}^n
\Bigl[{\cal N}_j {\cal A} + \zeta {\cal B} \Bigr],
\label{dirprod}
\ee
where the number of end branes is $\zeta =0,1$ or $2$ depending on
the boundary conditions.
It is then obvious that the number of elementary cells
which end on the $AdS$ horizon in $p$ directions is
given by the coefficient of ${\cal B}^p$
in the expansion of (\ref{dirprod}):
\be
\nu_p = \zeta^n \sum_{{\cal C}_p} \prod_{\alpha \in {\cal C}_p}
[\frac{{\cal N}_{\alpha}}{\zeta}],
\ee
where ${\cal C}_p$ is any set of $n-p$
distinct elements from the set $\{1, ..., n\}$, and
$[r]$ denotes rounding off the number $r$ to the nearest smaller
integer.
Specifically, the number of
cells inside the crystal, with $p=0$ is
$\nu_0 = \zeta^n \prod_{k=1}^n [\frac{{\cal N}_k}{\zeta}]$.
Note that the total number
of branes along $k^{\rm th}$ direction is set by the boundary
conditions $(i)$-$(iv)$.
Hence we obtain the following
formula for the $4D$ Planck mass:
\be
M^2_{Pl} = \frac{\zeta^n M^{n+2}_*}{(n+1)! K^n} \sum_{p=0}^n
\sum_{{\cal C}_p} \prod_{\alpha \in {\cal C}_p} [\frac{{\cal
N}_{\alpha}}{\zeta}]
\sum_{j=0}^{n-p} (-1)^j
\sum_{{\cal B}_j}
\frac{1}{(1+ K \sum_{i \in {\cal B}_j} l_i)^2},
\label{mpl}
\ee
where ${\cal B}_j$ are combinations of length $j$ of elements of
$\{l_{c(1)},...,l_{c(n-p)}\}$, and $c(k) \in {\cal C}_p$.
In the special case of finite crystals with cubic global symmetry
and two end branes,
there is the same number of branes in each direction,
${\cal N}_1 = ... = {\cal N}_n = {\cal N}$, and the lattice
spacing is the same in all directions, $l_1 = ... = l_n = l$.
Therefore all ${\cal C}_p$'s of the same length coincide,
the symmetry factors $\nu_p$ are
${\nu}_p = 2^n [\frac{{\cal N}}{2}]^{n-p} \frac{n!}{p!(n-p)!}$, and hence
\be
M^2_{Pl} = \frac{2^n M^{n+2}_*}{(n+1) K^n} \sum_{p=0}^n
\sum_{j=0}^{n-p} \frac{(-1)^j}{j!p!(n-p-j)!}
\frac{1}{(1+jKl)^2} [\frac{{\cal N}}{2}]^{n-p}.
\label{mplspec}
\ee
It is now obvious that for either class of finite crystals, $(ii)$
or $(iv)$, the formula (\ref{mpl}) gives a finite value for the
$4D$ Planck mass, meaning that there is a $4D$ graviton. In
effect, those crystal configurations effectively ``compactify" the
extra dimensions. In the case of infinite crystals $(i)$ and $(iii)$,
although each individual contribution to (\ref{mpl}) is finite,
formally the sum diverges since there is an infinite number of
cells to sum over. However, the sum can be renormalized at the
expense of loosing a one-to-one correspondence between the $4D$
and $(4+n)D$ Planck scales. Indeed, for infinite crystals
$M^2_{Pl} \sim {\cal N}^n M^{2+n}_* L^n \rightarrow \infty$, since
${\cal N} \rightarrow \infty$. Since the divergence comes
from the bulk volume being infinite, the
renormalization must be a bulk phenomenon.
To renormalize $M_{Pl}$, we recall
that since by the fine-tuning condition (\ref{finetun}) the bulk
cosmological constant is related to the brane tensions, and since
the brane tensions must be unaffected by a bulk renormalization,
the $AdS$ radius must also be unaffected. Hence, the only way to
remove infinity from $M_{Pl}$ is to take $M_*$ to be infinite, and
renormalize it by $M_{\rm bare} \rightarrow {\cal N}^{-n/(n+2)}
M_*$. But this means that the resulting mass $M_*$ need not
be related to the scale of the brane physics, and hence this
would seem to require another fine tuning to explain the hierarchy between
$M_{Pl}$ and $m_{EW}$.

On the other hand, eq. (\ref{mplspec}) shows that for finite
crystals there are additional possibilities for generating
hierarchies. This is because $M_{Pl}$ depends on four different scales:
the fundamental scale $M_*$, the $AdS$ radius
$L = (\sqrt{n} K)^{-1}$, the lattice spacing $l$ and the total
volume of the crystal $\sim ({\cal N}l)^n$.
Hence it is easier to generate exponential hierarchies in large
crystals, even if all the other scales are of the same order of
magnitude.

Before we move on to study the bulk gravitons, we should note
that infinite and semi-infinite cases $(i)$ and $(iii)$
are also disfavored phenomenologically, even if we perform a renormalization
as outlined above. In an infinite crystal there is
an infinite number of intersections. Discrete translational invariance
of the crystal would require that the energy density of particles on
each intersection is of the same order (which would remain true
even if the symmetry is weakly broken).
Because at long distances the gravitational force is dominated by
the zero modes, an observer on any intersection would feel the
gravitational field of particles localized to all the
intersections in the crystal regardless of how far in the bulk
they are. Thus the localized energy density everywhere would add up
to give the total energy density controlling the cosmological evolution of
all intersections. Hence to sum up to a finite value,
the energy density on each intersection must be infinitesimal.
This implies that all but an
infinitesimal amount of the energy density of the Universe
we experience would have to be dark.
Note that this remains true {\it regardless} of any
renormalization of the scales, which may or may not be necessary
(in a fashion similar to that for $M_{Pl}$). The matter on distant
intersections would still be dark, there would be infinitely
many of them, and renormalization would not change the ratio
of dark matter to visible matter.
Since the amount of visible matter in our Universe is not infinitesimal,
we see that at least one intersection in the crystal must have
finite $\rho$, and most of the others cannot.
Hence the discrete translational invariance
of the crystal must be strongly broken.
Our intersection is a defect of the lattice. It cannot deform the lattice
too strongly, in order not to loose the $4D$ graviton. But
it does suggest that the graviton zero mode accumulates around
the defect, which deforms the background more than most of the other
vertices in the lattice. This can generate a peaked profile for the
wave function
of the zero mode, making other intersections weakly coupled. Such
possibilities for generating asymmetric mirror worlds have been
discussed in \cite{addk2}.
In practice, therefore, an infinite crystal would have to be
physically hardly distinguishable from a finite crystal,
with the exception that the strong
coupling regime, discussed in \cite{rs}, could be absent.
A finite crystal would have to be consistent with observational
constraints much like the Manyfold universes in theories with
sub-millimeter internal dimensions \cite{addk2}.
Thus, an infinite crystal with unbroken discrete translational
invariance may be only a mathematical idealization. We will retain
it, however, since it is quite useful for the study of bulk
gravitons, to which we now turn.

The field equation for linear perturbations
around the background solution can be obtained by expanding the
$4D$ part of the
metric according to $\bar g_{\mu\nu} = g_{\mu\nu} + h_{\mu\nu}$,
where $h_{\mu\nu}$ is in the transverse traceless gauge with \
respect to the background $\nabla_\mu h^\mu{}_\nu = h^\mu{}_\mu =
0$, and expanding the
Einstein's equations to linear order in $h_{\mu\nu}$. The
perturbations satisfy
\be
\delta R_{\mu\nu} = \frac12 R^\lambda{}_\nu h_{\mu\lambda} +
\frac12 R^\lambda{}_\mu h_{\lambda\nu}.
\ee
After evaluating $\delta R_{\mu\nu}$, substituting the
background crystal metric (\ref{finsol}), and
defining the wave function $\Psi$ by
\be
h_{\mu\nu} = \Omega^{(2-n)/2} \Psi,
\label{wavefdef}
\ee
where $\Psi$ in general denotes a complex function, which
arises as a linear combination of $+$ and $\times$ polarizations,
we find
\ba
&&\frac12 \Box_4 \Psi + \frac12 \vec \nabla^2 \Psi
- \frac{n(n+2)(n+4)}{8 (\sum_k {\cal S}(z^k) + \sqrt{n}L)^2} \Psi
\nonumber \\
&&~~~~~~~~~~~~~~~~~~~~
+ \frac{n+2}{2(\sum_k {\cal S}(z^k) + \sqrt{n}L)} \sum_{k=1}^{n}
\sum_{j} (-1)^j \delta(z^k - j l_k) \Psi = 0.
\ea
In this equation, the $4D$ D'Alembertian is solved by the expansion into
plane waves, $\Box_4 \Psi = m^2 \Psi$, where $m$ is the KK mass.
The resulting equation is
\be
\vec \nabla^2 \Psi + \Bigl({m^2}
- \frac{n(n+2)(n+4)}{4 (\sum_k {\cal S}(z^k) + \sqrt{n}L)^2}
+ \frac{n+2}{(\sum_k {\cal S}(z^k) + \sqrt{n}L)} \sum_{k=1}^{n}
\sum_{j} (-1)^j \delta(z^k - j l_k) \Bigr) \Psi = 0.
\label{schro}
\ee
This equation is a static Schr\"odinger
equation for an electron in a periodic potential.
It is now clear that the lowest energy eigenfunction of (\ref{schro}),
which corresponds to the $4D$ graviton zero mode, is
$\Psi_0 \sim {\rm const}$ with $m^2=0$, which by (\ref{wavefdef})
reproduces precisely (\ref{zeromode}), as we claimed.
For $m^2 > 0$, the analogy with the electron in a crystal shows
that the solutions of (\ref{schro}) are Bloch waves, satisfying
$\Psi_{\vec q}(\vec z) = \exp(i\vec q \cdot \vec z) v_{\vec
q}(\vec z)$, where by Floquet's theorem $v_{\vec q}$ are periodic
functions under the action of the discrete translation group. The
energy spectrum has a band structure due to resonant scattering.
The bands are continuous in infinite crystals, and discrete
in finite ones. In addition, there is a mass gap, separating
the lowest-lying excited state from the ground state by a finite energy.
In the language of our crystal Manyfold, this means that the
lightest KK graviton must have a finite mass, and that the KK
modes come in bands. As a result, their influence on
the low energy physics on the intersection will be significantly
suppressed when compared to a single brane of \cite{rs} or
a single intersection of \cite{addk1}. This is true not only
for the production of bulk gravitons by the collisions of the Standard Model
particles, but also for the corrections to Newton's law of
gravity. Due to the gap, there will be an exponential suppression
of the higher-order corrections to the inverse square law.
Mass gap was also discussed in \cite{bras} and in \cite{nam}.

To be more concrete, we focus on
an infinite crystal with only one internal dimension.
Finite crystals will behave similarly, when they consist of many branes,
except that the allowed bands will be discrete rather than continuous.
We expect that the qualitative properties of crystals with more than
one internal dimension will remain similar.
In the case of one extra dimension, the Schr\"odinger equation
(\ref{schro}) reduces to
\be
\Psi''
+\Bigl(m^2- \frac{15}{4 ({\cal S}(z) + L)^2}\Bigr) \Psi
+ \frac{3}{{\cal S}(z) + L}
\sum_{j} (-1)^j \delta(z - j l) \Psi = 0.
\label{schone}
\ee
By the analogy with electrons in a crystal, e.g. with the
Kronig-Penney model, we need to solve this equation inside
two adjacent elementary cells in the crystal. By the periodicity
of (\ref{finsol}), we can choose the elementary cells $0<z<2l$ and
$2l<z<4l$.
The $\delta$-function potentials in (\ref{schone}) can be recast
as the boundary conditions on the derivatives of $\Psi$ at the
vertices: $\Psi'(l_+)-\Psi'(l_-) = \frac{3}{l+L}\Psi(l)$ and
$\Psi'(2l_+)-\Psi'(2l_-) = -\frac{3}{L}\Psi(2l)$. Moreover, the
wave function at each vertex must satisfy $\Psi(vertex_+) =
\Psi(vertex_-)$ in order to conserve probability.
The differential equation in the first elementary cell is
\be
\Psi'' + m^2 \Psi = \cases{\frac{15}{4(z+L)^2}  \Psi, & if $0 < z < l$; \cr
\frac{15}{4(2l-z+L)^2} \Psi, & ~~~$l< z < 2l $. }
\ee
These are Bessel equations. Using the boundary conditions above,
the solution is
\be
\Psi = \cases{\sqrt{m(z+L)} \Bigl[A H^+_2m(z+L))
+ B H^-_2(m(z+L)) \Bigr],
& $0 < z < l$; \cr
\sqrt{m(2l-z+L)} \Bigl[\Bigl(
\frac{h^+_1 h^-_2 + h^-_1 h^+_2}{ h^-_1 h^+_2-h^+_1 h^-_2 } A
+ \frac{2 h^-_1 h^-_2}{ h^-_1
h^+_2-h^+_1 h^-_2} B \Bigr) H^+_2m(2l-z+L))  \cr
~~~~~~~~~~~~~~~~ + \Bigl(
\frac{2  h^+_1 h^+_2}{h^+_1 h^-_2 - h^-_1 h^+_2} A +
\frac{h^+_1 h^-_2 + h^-_1 h^+_2}{h^+_1 h^-_2 - h^-_1
h^+_2} B \Bigr)
H^-_2(m(2l-z+L)) \Bigr], & $l< z < 2l $.}
\label{wavsol}
\ee
Here $H_2^{\pm}$ are the Hankel functions, which generalize the
plane waves for the case of the Bessel equation. They should be
used instead of the real Bessel functions, since they are needed
to encode the momentum by phase rotation.
It is rather amusing that both {\it classical}
$4D$ graviton polarizations are needed to mimic the quantum-mechanical
phase. Bloch waves physically correspond to rotation of
the polarization plane of bulk gravitons in the
crystal. In evaluating the transmission and reflection
coefficients in (\ref{wavsol}) on the $\delta$-function, we have
used a recursive relation $\frac{dH_2^{\pm}}{dw} = H^\pm_1 - \frac{2}{w}
H_2^\pm $. Then we can evaluate the functions and derivatives at $l_\pm$,
and using the boundary conditions eliminate $C$ and $D$ in
favor of $A$ and $B$, and have the function in each elementary
cell be determined by two (complex) constants, just like in
the Kronig-Penney model.

The solution in the adjacent cell can be obtained similarly.
It is given by
\be
\Psi = \cases{\sqrt{m(z-2l+L)} \Bigl[\hat A H^+_2m(z-2l+L))
+ \hat B H^-_2(m(z-2l+L)) \Bigr],
& $2l < z < 3l$; \cr
\sqrt{m(4l-z+L)} \Bigl[\Bigl(
\frac{h^+_1 h^-_2 + h^-_1 h^+_2}{ h^-_1 h^+_2-h^+_1 h^-_2 } \hat A
+ \frac{2 h^-_1 h^-_2}{ h^-_1
h^+_2-h^+_1 h^-_2} \hat B \Bigr) H^+_2m(4l-z+L))  \cr
~~~~~~~~~~~~~~~~ + \Bigl(
\frac{2  h^+_1 h^+_2}{h^+_1 h^-_2 - h^-_1 h^+_2} \hat A +
\frac{h^+_1 h^-_2 + h^-_1 h^+_2}{h^+_1 h^-_2 - h^-_1
h^+_2} \hat B \Bigr)
H^-_2(m(4l-z+L)) \Bigr], & $3l< z < 4l $.}
\ee
Now we need to relate $A,B,\hat A, \hat B$. The boundary
conditions which they must satisfy are
\ba
&&\Psi(2l) = \exp(2iql) \Psi(0) ,\nonumber \\
&&\Psi(4l) = \exp(2iql) \Psi(2l) ,\nonumber \\
&&\Psi(2l_+) = \Psi(2l_-) ,\nonumber \\
&&\Psi'(2l_+) - \Psi'(2l_-) = - \frac{3}{L} \Psi(2l) ,
\ea
where the former two come from the Floquet's
theorem for the Bloch waves and the
periodicity of the braneworld crystal,
and the latter two come from the continuity of $\Psi$ and the jump
of $\Psi'$ at $z=2l$. The parameter $q$ represents
the wave vector of the Bloch wave, which is to be determined.
The first three conditions yield
\ba
B &=& \frac{(h^+_1 h^-_2 + h^-_1 h^+_2)\hat h^+_2
- (h^-_1 h^+_2 - h^+_1 h^-_2)
\exp(2iql) \hat h^+_2
- 2 h^+_1 h^+_2 \hat h^-_2 }{(h^+_1 h^-_2 + h^-_1 h^+_2)\hat h^-_2
+ (h^-_1 h^+_2 - h^+_1 h^-_2)
\exp(2iql)\hat h_2^- - 2 h^-_1 h^-_2 \hat h_2^+} A , \nonumber \\
\hat B &=& \frac{(h^+_1 h^-_2 + h^-_1 h^+_2)\hat h^+_2
- (h^-_1 h^+_2 - h^+_1 h^-_2)
\exp(2iql) \hat h^+_2
- 2 h^+_1 h^+_2 \hat h^-_2 }{(h^+_1 h^-_2 + h^-_1 h^+_2)\hat h^-_2
+ (h^-_1 h^+_2 - h^+_1 h^-_2)
\exp(2iql)\hat h_2^- - 2 h^-_1 h^-_2 \hat h_2^+} \hat A ,\nonumber \\
\hat A &=& \exp(2iql) A ,
\ea
where we have defined the numbers $h_k^\pm = H^\pm_k(m(l+L))$ and
$\hat h_k^\pm = H^\pm_k(mL)$.
The last condition will give an equation relating $q$ and $m$,
because the normalization constant $A$ factors out: it is determined by
the overall normalization of the wave function.
The equation for $q$ is the band equation. A straightforward
computation gives
\be
\cos(lq) = \frac{
(j_2n_1+j_1n_2)(\hat j_2 \hat n_1+ \hat j_1 \hat n_2)
- \hat j_1 \hat j_2 (j_1 j_2+3n_1n_2)
- \hat n_1 \hat n_2 (3j_1 j_2+n_1n_2)}{2(j_2 n_1 - j_1 n_2)
(\hat j_2 \hat n_1 - \hat j_1 \hat n_2)} .
\label{bandeq}
\ee
In this equation, we have decomposed the Hankel
functions into the Bessel functions $J_k$ and $N_k$,
$J_k = {\rm Re}~ H^+_k, N_k = {\rm Im}~ H^+_k$,
with the definitions $j_k = J_k(m(l+L))$, $n_k=N_k(m(l+L))$,
$\hat j_k = \hat J_k(mL)$ and $\hat n_k=\hat N_k(mL)$.
To see that there is a gap in the spectrum already at zero
momentum, we can substitute $q=0$ in (\ref{bandeq}). The resulting
condition for the lowest-lying mass $m$ is the gap equation:
\be
j_1 j_2 \hat j_1 \hat j_2 + n_1 n_2 \hat n_1 \hat n_2
+ j_2 \hat j_2 n_1 \hat n_1 + j_1 \hat j_1 n_2 \hat n_2 =
3 j_1 \hat j_2 \hat n_1 n_2 + 3 \hat j_1 j_2 n_1 \hat n_2
- 3 \hat j_1 \hat j_2 n_1 n_2 - 3 j_1 j_2 \hat n_1 \hat n_2.
\label{gap}
\ee
It is evident that this equation provides a non-trivial
constraint on $m$. Finding the lowest positive value of $m$
which solves this equation is
a straightforward, albeit tedious task.
However, for $l>L$ we can readily obtain the correct order of magnitude
of the gap. Since the Bessel functions in (\ref{gap}) are
evaluated at $m(l+L)$ and $mL$, the only two candidates for
mass scales for $m_{\rm gap}$
are $L^{-1}$ and $(l+L)^{-1}$. On the other hand, in the limit
$l \rightarrow \infty$, the crystal solution reduces to the single
brane solution of \cite{rs}, where the bulk gravitons do not have
a gap. This excludes $L^{-1}$. Hence the mass gap must be
\be
m_{\rm gap} = \frac{{\cal O}(1)}{l+L}.
\label{gapsol}
\ee
With this at hand, it is easy to estimate the corrections to
Newton's law in $4D$ induced by bulk gravitons. By the form of the
wave functions (\ref{wavsol}), the $4D$ Newtonian potential for particles
localized to a $3$-brane at the intersection of $4$-branes
with positive tension (``Planck\footnote{The nickname ``Planck" refers to the
$5D$ Planck scale $M_*$, not the $4D$ $M_{Pl}$.}  brane"), with the
corrections from the bulk gravitons, is
\be
V_N = - G_N \frac{M_1 M_2}{r} \Bigl(1
+ {\rm a} \int_{m_{\rm gap}}^\infty \frac{dm}{K} \,
\frac{m}{K} e^{-mr} \Bigr) ,
\ee
which upon integration gives
\be
V_N \simeq - G_N \frac{M_1 M_2}{r} \Bigl(1+ {\rm a}
\frac{L^2}{(L+l)r} e^{-{\rm b} r/(L+l)}
\Bigr) ,
\label{newpotplus}
\ee
where ${\rm a}$ and ${\rm b}$ are constants of order unity.
Note that because of the gap, the corrections are
${\cal O}(r^{-2})$, in contrast to ${\cal O}(r^{-3})$ in \cite{rs}.
On the other hand, using eq. (\ref{finsol})
when $l > L$, the potential for particles localized
to a $3$-brane at the intersection of branes with negative tension
(``$TeV$ brane") is, roughly,
\be
V_N = - G_N \frac{M_1 M_2}{r} \Bigl(1
+ {\rm a} \int_{m_{\rm gap}}^\infty \frac{dm}{K} \,
\Bigl(\frac{l}{L}\Bigr)^3 e^{-mr} \Bigr) ,
\ee
which gives
\be
V_N \simeq - G_N \frac{M_1 M_2}{r} \Bigl(1+ {\rm a}
\frac{l^3}{L^2 r} e^{-{\rm b} r/l}
\Bigr) ,
\label{newpotminus}
\ee
and again  ${\rm a}$ and ${\rm b}$ are (different) constants of order unity.
Clearly the corrections to Newton's law for particles on the
Planck brane can remain small at
sub-millimeter distances even if $L \gg l_{Pl} \sim 10^{-34} mm$,
despite the fact that there is only five spacetime dimensions.
However, the corrections become more significant on the $TeV$
brane, as is clear from (\ref{newpotminus}). But if we are to
solve the hierarchy problem while suppressing the corrections to $4D$
gravity to satisfy the observable bounds, our world should be
a $TeV$ brane. Therefore, the existing
observational constraints yield $l^3 \le 1 mm \cdot L^2$, i.e.
$l \le 1 mm \cdot (\frac{L}{l})^2 < 1 mm$.
Further, if $l > L$, eq. (\ref{mplspec}) gives
$M^2_{Pl} \sim M^3_* L {\cal N}$, and so the $4D$ Planck mass
depends on the size of the crystal and the lattice spacing.
The fundamental scale should satisfy
$M_* \sim \frac{l}{L} \cdot TeV$. Hence $M_{Pl}^2 \sim (TeV)^3
~\frac{l^3}{L^2}~ {\cal N}$,
and using $l \sim eV^{-1} \ll 1 mm$, for which
$L \sim 10^{12} M^{-1}_*$ ensuring the validity of the
supergravity description of the solution (\ref{finsol}), gives
${\cal N} \sim 10^{16}$. With this choice of numbers, the
Standard Model on the $TeV$ branes
has scale $m_{EW} \sim 1 TeV$, and its mirror on the
Planck branes has scale $M_* \sim 100 TeV$. While the
number of branes required is clearly very large,
being equal to $\frac{M_{Pl}}{TeV}$, i.e.
to the hierarchy between the scales,
it is smaller by sixteen orders of magnitude than the maximal number
of branes which fit inside the flat extra
dimensions \cite{admr}.
Note also that the size of the crystal is ${\cal N} l \sim 10^{12} mm$,
with the choice $M_* \sim 100 TeV$, which is four orders of
magnitude smaller than the size of a flat compact dimension
required to generate the same ratio $\frac{M_{Pl}}{m_{EW}}$
\cite{add}. The required number of
branes could be reduced in the case of more extra
dimensions, but the details are beyond the scope of the present
article.

It is straightforward to generalize this discussion when the lattice
spacing varies within each family of parallel branes. In the extreme
case, the configuration may be more similar to few separate crystals
residing at different places in the $AdS$ bulk. The effective
scale would change from crystal to crystal, and in the infrared
limit the low energy $4D$ physics on distant crystals would be
similar to that on a $3$-brane far from an intersection \cite{addk1},
or far from the Planck brane in $5D$ \cite{lr},
or to a Manyfold with bulk profiles \cite{addk2}. With several
crystals in the bulk, it is easier to generate hierarchy with
fewer branes in each crystal, as is straightforward to verify.
In fact, separated ``small crystals" would correspond precisely
to the cases of distant branes discussed in \cite{addk1,lr}, where
hierarchy is generated by gravitational shining.

We close with several concluding remarks. We have constructed general
solutions describing intersecting families of parallel $n+2$-branes in
$4+n$-dimensional $AdS$ space. The families intersect at right
angles, and consist of branes with both positive and negative
tension. As a result, there arise several different types of
crystals, both finite and infinite. All of them give rise to the
usual $4D$ gravity, but phenomenological
considerations strongly favor either finite crystals or infinite ones
with broken translational invariance, profiles in the bulk and our
world as a lattice defect. If such solutions are interlaced
by $3$-branes, at low energies they behave the same as Manyfold
universes of \cite{addk2}. The periodicity of the crystal lattice
implies that the bulk gravitons are Bloch waves, with a band
spectrum and a gap at zero energy. We construct the spectrum
explicitly in the case of one extra dimension, and find that in
the extreme case it may be possible to have the string scale at
$\sim 100 TeV$, if there is $\sim 10^{16}$ branes in the
configuration. The crystal worlds raise many interesting questions:
the issues of stability, early cosmology, etc. While the bulk gravitons
behave as electrons in the usual crystal, one may wonder which
degrees of freedom play role of the conventional phonons. It seems
plausible that they should be the Goldstone modes associated with
the branes in the crystal, which may get excited by sufficiently
strong external perturbations. A more precise consideration would
be needed to test this. Further, it would be interesting to find
the properties of
higher-dimensional crystals, since they may require fewer branes
to resolve the hierarchy problem while ensuring Newton's law of
gravity in $4D$ at distances larger than a millimeter. We hope to
return to some of these issues in the future.


\section*{Acknowledgments}

The author acknowledges valuable discussions with Savas Dimopoulos
and John March-Russell. This work has been supported in part
by NSF Grant PHY-9870115.


\end{document}